\documentclass[11pt]{article}
\usepackage{graphicx,amssymb}
\usepackage{epsfig}
\usepackage{graphics}
\usepackage{psfrag}
\newcommand{\ba}{\begin{array}}
\newcommand{\ea}{\end{array}}
\newcommand{\bd}{\begin{displaymath}}
\newcommand{\ed}{\end{displaymath}}
\newcommand{\be}{\begin{equation}}
\newcommand{\ee}{\end{equation}}
\newcommand{\bea}{\begin{eqnarray}}
\newcommand{\eea}{\end{eqnarray}}

\newcommand{\rpv}{\mbox{$\not \hspace{-0.10cm} R_p$ }}

\def\beq{\begin{equation}}

\def\eeq{\end{equation}}

\def\bea{\begin{eqnarray}}

\def\eea{\end{eqnarray}}

\def\bq{\begin{quote}}

\def\eq{\end{quote}}

\def\Journal#1#2#3#4{{#1} {#2} (#4) #3}

\def\NPB{Nucl. Phys. B}

\def\PLB{Phys. Lett. B}

\def\PRD{Phys. Rev. D}

\def\PR{Phys. Rept.}

\begin{document}

\vspace*{-0.5in}

\renewcommand{\thefootnote}{\fnsymbol{footnote}}

\begin{flushright}

LPT-ORSAY-09-99 \\

\end{flushright}

\vskip 5pt

\begin{center}
{\Large {\bf Dark matter and neutrino masses in the R-parity violating NMSSM}} \vskip 25pt {\bf C.-C. Jean-Louis
$^{1,}$\footnote{E-mail
 address: charles.jean-louis@th.u-psud.fr}},
{\bf G. Moreau $^{1,}$\footnote{E-mail
 address: gregory.moreau@th.u-psud.fr}}
\vskip 10pt $^1${\it Laboratoire de Physique Th\'eorique,
Universit\'e
de Paris-sud XI} \\
{\it B\^at. 210, 91405 Orsay, France} \\
\normalsize
\end{center}

\begin{abstract}
The R-Parity symmetry Violating (RPV) version of the Next-to-Minimal Supersymmetric Standard Model (NMSSM)
is attractive simultaneously with regard to the so-called $\mu$-problem
and the accommodation of three-flavor neutrino data at tree level. In this context, we show here that if the Lightest Supersymmetric Particle (LSP) is the gravitino,
it possesses a lifetime larger than the age of the universe
since its RPV induced decay channels are suppressed by the weak gravitational strength. This conclusion holds
if one considers gravitino masses $\sim 10^2$ GeV like in supergravity scenarios, and is robust if the lightest pseudoscalar Higgs field is as light as $\sim 10$ GeV
[as may occur in the NMSSM]. For these models predicting in particular an RPV neutrino-photino mixing,
the gravitino lifetime exceeds the age of the universe by two orders of magnitude.
However, we find that the gravitino cannot constitute a viable dark matter candidate since its too large RPV decay widths
would then conflict with the flux data of last indirect detection experiments.
The cases of a sneutrino LSP or a neutralino LSP as well as the more promising
gauge-mediated supersymmetry breaking scenario are also discussed. Both the one-flavor
simplification hypothesis and the realistic scenario of three neutrino flavors are analyzed.
We have modified the NMHDECAY program to extend the neutralino mass matrix to the present framework.
\end{abstract}

\newpage

\renewcommand{\thesection}{\Roman{section}}
\setcounter{footnote}{0}
\renewcommand{\thefootnote}{\arabic{footnote}}

\section{Introduction}
\label{intro}

In supersymmetry, the superpartner of the graviton, namely the
so--called gravitino, plays a central theoretical role as it
constitutes the gauge fermion of supergravity theories
\cite{SUGRA,SUGRAreport}. From the cosmological point of view, the gravitino
may be the Lightest Supersymmetric Particle (LSP), depending on
the way supersymmetry is broken, and hence constitute a stable
dark matter candidate \cite{GravDM}.
Interestingly, one could even assume supersymmetric models where
the R--parity symmetry \cite{rpar1} is violated. Then the
gravitino becomes unstable due to new decay channels into Standard
Model (SM) particles induced by the R--Parity Violating
(RPV) interactions. However, because of the small gravitational
interaction and possibly weak RPV couplings, the gravitino lifetime
can still exceed the age of the universe (by several orders of
magnitude) in the particular case of RPV bilinear terms \cite{Takayama}
remaining thus a realistic dark matter
candidate. This scenario can yield a complete cosmological
framework also incorporating leptogenesis and nucleosynthesis, as
recently shown in Ref.~\cite{NucleoLepto}. In such a scenario, the
gravitino decays into SM particles lead to specific signatures in
high--energy cosmic rays; the produced flux of gamma rays and
positrons can even account \cite{Koji} respectively for the extragalactic
component of the excess in the HEAT \cite{HEAT} data. The decaying
gravitinos could also constitute an interpretation (see below) of the exotic
positron source recently discovered by the PAMELA Collaboration \cite{PAMELA}.

Another advantage of the existence of RPV interactions is the
induced mixing between left--handed neutrinos and neutral
gauginos. This mixing mechanism generates neutrino masses
economically, without extending the SM field content to additional
right--handed neutrino fields. There exists a tension between
generating a sufficiently large neutrino Majorana mass scale
[requiring strong RPV] and, at the same time, keeping the
gravitino lifetime larger than the age of the universe [imposing
weak RPV]. Nevertheless, this tension leaves some acceptable
windows in the parameter space
\cite{Takayama} for example with gravitino masses around $100$ GeV
in the Minimal Supersymmetric Standard Model (MSSM).

In the present paper, we re-consider the above scenario in the
framework of the Next-to-Minimal Supersymmetric Standard Model
(NMSSM) \cite{NMSSM} (see recent reviews \cite{NMpheno} for the phenomenological
studies). An increasingly important virtue of the NMSSM is that it
improves the `little hierarchy' problem originating from the
requirement of large soft supersymmetry breaking masses compared
to the ElectroWeak (EW) scale. It also provides a solution to the
so-called $\mu$-problem by arranging the vacuum expectation value
(vev) of a new gauge singlet scalar field of order of the
supersymmetry breaking scale, so that the $\mu$ parameter turns
out to be at the EW scale. We have shown in a previous work
\cite{AGG} that in the NMSSM with bilinear RPV terms, two
non-degenerate massive neutrino states can emerge at tree level,
in contrast with the MSSM case where only one neutrino eigenstate
acquires a mass (at tree level). Hence, the global three-flavor
neutrino data can be accommodated \cite{Valle}, at tree level,
without appealing to loop dynamics which is vulnerable to
model-dependent uncertainties.

In such a framework, both the gaugino mass matrix and RPV mixing
terms are modified w.r.t. the MSSM case. Furthermore, the NMSSM provides the possibility
\footnote{Such a possibility was in particular motivated by interpreting the well-known LEP excess at $2.3 \sigma$ in the $Z + 2b$ channel -- via the production
of a light Higgs boson ($m_h \simeq  99$ GeV) \cite{Gunion} -- but this NMSSM interpretation seems to be excluded by a recent ALEPH analysis \cite{ALEPH}.}
of reducing significantly w.r.t. MSSM the mass of the lightest pseudoscalar Higgs field [which can
constitute partially the gravitino decay final state]. Moreover, the specific gauge singlet scalar
field gives rise to new contributions to the gravitino RPV decay. Hence, the double
question in the NMSSM on the possibility of generating the correct neutrino mass scale, while
still keeping the gravitino as a good dark matter candidate, is relevant and well motivated.
\\ We find in this paper that the answer concerning the gravitino stability can be positive for parameters
passing the theoretical and phenomenological constraints
implemented in the NMHDECAY code \cite{NMHDECAY} and in particular
for gravitino masses as large as $m_{\rm 3/2} ={\cal O}(10^2)$ GeV [tending to increase the total gravitino width].
This is satisfactory as it corresponds to the typical scale of supersymmetry breaking in supergravity scenarios.
In both scenarios where the supersymmetry breaking is mediated purely by gravity \cite{SUGRA,SUGRAreport}
or partially by gravity and gauge interactions \cite{hybridI} (such classes of hybrid models have been
recently motivated in string inspired constructions \cite{hybridII}),
the gravitino is the LSP in wide regions of the parameter space and its typical mass is
in the range between $100$ GeV and $1$ TeV.
Besides, thermal leptogenesis and universal boundary conditions
for gaugino masses at the Grand Unification Theory (GUT) scale
restrict $m_{\rm 3/2}<600$ GeV \cite{thermal} and, on the other
hand, the interpretation of the PAMELA positron anomaly imposes
$m_{\rm 3/2} \gtrsim 200$ GeV \cite{Buch}.
\\ Nevertheless, our second result on a gravitino LSP around $\sim 10^2$ GeV is that its RPV decay rates
lead to fluxes exceeding the total flux measured in last indirect detection experiments.
\\ We will finally discuss the pure case, still within the NMSSM, of Gauge-Mediated Supersymmetry Breaking (GMSB)
where the LSP gravitino mass verifies typically $m_{\rm 3/2} \gtrsim 1$ eV \cite{GMSB} and the neutralino mass matrix receives
some modifications \cite{nmssmtools3}. Conclusions here are more optimistic.

Let us make some comments on the results obtained
in the present paper. First, we have considered all the possible
gravitino decays, namely into photon, Z, W bosons and
(pseudo)scalar Higgs fields, in contrast with Ref.~\cite{Takayama}
where only the photon channel was considered. The new
kinematically allowed channels open up due to the higher
gravitino masses considered here.
Secondly, the trilinear RPV couplings are also discussed.
Furthermore, we investigate for the first
time the realistic case of three flavors of neutrinos. We
conclude that the complete RPV mixing obtained in this case does
not invalidate the positive results obtained for the simultaneous
solution of the neutrino mass and dark matter problems. Finally,
the pure neutrino results include the various NMSSM-constraints
implemented in NMHDECAY \cite{NMHDECAY}
(that we have modified to include the neutrino components in the neutralino mass matrix)
and are derived from a numerical treatment of the full neutral gaugino mass matrix, in
contrast with preliminary work in Ref.~\cite{AGG}.
The modified version of NMHDECAY thus contains an implementation of the whole neutralino mass matrix
which includes, in the present context, some matrix elements induced by the RPV couplings and responsible for
the mixing between the higgsinos and neutrinos (see Section \ref{NeutMassMat}).

In addition, we will explore the alternative possibilities of the
neutralino and sneutrino as stable LSP dark matter candidates,
under the same assumption of the desired neutrino mass spectrum
generated through the RPV mixing in the NMSSM. The philosophy is
to establish a systematic list of the viable supersymmetric dark
matter candidates. Since this topic relies on the particle
spectrum, we base our study on a systematic exploration of the
parameter space. Our conclusions are also based on new
calculations of the neutralino and sneutrino decay channels.

Let us mention previous related works.
The neutrino flux from direct gravitino decays has been computed
in a simple scenario with bilinear R-parity breaking \cite{Flux}. The
diffuse gamma ray flux was also studied in Ref.~\cite{FluxGamma}.
A realization in minimal supersymmetric left-right models
within supergravity, with a gravitino LSP dark matter and R-parity breaking,
was proposed in Ref.~\cite{Moha} (see also Ref.~\cite{Pavel}). Finally,
a scenario with R-parity violation in the right-handed neutrino sector
was motivated by the PAMELA data \cite{Endo}.
\\ Concerning supersymmetric models for neutrinos -- independently of the dark
matter problem -- RPV versions of the NMSSM have been previously studied in
Ref.~\cite{rpNMSSM0}, Ref.~\cite{rpNMSSM1} and Ref.~\cite{rpNMSSM2}.
Besides, alternative supersymmetric extensions with broken $R$-symmetries have been proposed in
order to address simultaneously the
$\mu$-problem and the neutrino mass aspect \cite{SRoy,Munoz,Nilles}
(and see Ref.~\cite{MunozBis} for a gravitino dark matter decay study in this context).
\\ RPV supersymmetric scenarios have also been studied within the context of
neutrino astrophysics (see e.g. \cite{Dreiner,Gava}).
Finally, for studies of gravitino dark matter without RPV couplings,
see for instance Ref.~\cite{Moultaka} for the constrained MSSM or
GMSB scenarios.

The paper is organized as follows. In Section II, we elaborate the RPV scenario
and describe the corresponding neutralino mass matrix. In the following sections,
we discuss subsequently the cases of a sneutrino LSP (Section III), a neutralino LSP (IV)
and a gravitino LSP (V). Finally, we conclude in Section VI.

\section{The RPV version of the NMSSM}
\label{Considerations}

\subsection{Theoretical framework}
\label{Formalism}

We consider the NMSSM which possesses a superpotential containing
two dimensionless couplings $\lambda$ and $\kappa$ in addition to
the usual Yukawa couplings:
\begin{equation}
W_{\rm NMSSM}=
Y^u_{ij} Q_i H_u U_j^c + Y^d_{ij} Q_i H_d D_j^c + Y^\ell_{ij} L_i H_d E_j^c
+ \lambda S H_u H_d + \frac{1}{3} \kappa S^3 \,
\label{WNMSSM}
\end{equation}
where $Y^{u,d,\ell}_{ij}$ are the Yukawa coupling constants
($i,j,k$ are family indexes), and $Q_i$, $L_i$, $U^c_i$, $D^c_i$,
$E^c_i$, $H_u$, $H_d$, $S$ respectively are the superfields for
the quark doublets, lepton doublets, up-type anti-quarks,
down-type anti-quarks, anti-leptons, up Higgs, down Higgs, extra
singlet under the standard model gauge group. An effective $\mu$
term, given by $\lambda \langle s \rangle H_u H_d$, is generated
via the vev of the scalar component $s$ ($\langle s \rangle$) of
the singlet superfield $S$.
\\ In case of GMSB \cite{nmssmtools3}, there are additional terms that must be added to the
NMSSM superpotential (\ref{WNMSSM}). Those terms are given by:
\begin{equation}
W_{\rm GMSB} \ = \ \xi_F S + \mu' S^2 \,
\label{GMSB}
\end{equation}
where $\xi_F$ ($\mu' $) is a new dimension-two (-one) parameter. These parameters in
GMSB models are generated at low energy as $S$ is coupled to the messenger sector.

Recalling that there exists no deep theoretical principle in
supersymmetry for the existence of an exact $R$-parity symmetry
\cite{PhysRep}, we adopt a generic approach by introducing both
the bilinear and trilinear RPV terms characteristic of the NMSSM
[the usual trilinear RPV couplings of the MSSM are considered in next section]:
\begin{equation}
W = W_{\rm NMSSM} + \mu_i L_i H_u + \lambda_i S L_i H_u ,
\label{Wgeneric}
\end{equation}
where $\mu_i$ ($\lambda_i$) are the dimension-one (dimensionless)
RPV parameters.

Within the usual NMSSM, only trilinear couplings with
dimensionless parameters (like $\lambda$ and $\kappa$) are kept in
the superpotential, while dimensional parameters (like $\mu$) are
generated from the vev $\langle s \rangle$. Here the RPV NMSSM
superpotential (\ref{Wgeneric}), containing a $\mu_i L_i H_u$
but no $\mu H_u H_d$ term, is assumed to arise in one of the
scenarios proposed in \cite{AGG}.

\subsection{Discussion on the trilinear RPV terms}
\label{Trilinear}

The most general RPV NMSSM superpotential also includes the
renormalizable trilinear RPV interactions:
\begin{equation}
W_{\rm RPV} = W +
\lambda_{ijk} L_i L_j E_k^c +
\lambda'_{ijk} L_i Q_j D_k^c +
\lambda''_{ijk} U^c_i D^c_j D_k^c.
\label{WgenericTRI}
\end{equation}

It is remarkable that if the order of magnitude for
$\lambda_{ijk}$ and $\lambda'_{ijk}$ is comparable to the values
of the Yukawa coupling constants for the electron and the down quark
[$Y^e$, $Y^d$], the exchange of leptons/sleptons and respectively
quarks/squarks in one-loop processes can generate Majorana
neutrino masses \cite{nubounds} in agreement with (or few orders
below) the oscillation experiment results: $10^{-3} \mbox{eV}
\lesssim m_{\nu_i} \lesssim 1 \mbox{eV}$.
This occurs for sfermion masses in the vicinity of $10^2$ GeV, the order of
supersymmetry breaking scale amount required from the gauge
hierarchy solution. Now, given the analog structure of the Yukawa
interactions (\ref{WNMSSM}) and the trilinear RPV terms
(\ref{WgenericTRI}), one can assume that the same
(flavor-)structure responsible of the $Y^{e,d}$ suppression
(w.r.t. $Y^t$ for the top quark) would also characterize
the $\lambda^{( \prime )}_{ijk}$ couplings, inducing
in turn the wanted tiny neutrino mass scale through one-loop
processes.
\\ Assuming such low values for the $\lambda^{( \prime )}_{ijk}$
coupling constants (which in general easily pass the various
phenomenological constraints \cite{REVbounds}), one finds from
formulas obtained in \cite{GravTri} that a gravitino ($\tilde G$)
LSP would decay into the three SM fermion final state $f_if_jf_k$
via the $\lambda^{( \prime )}_{ijk}$ interactions with a
sufficiently small width. Quantitatively, for a gravitino mass
around $m_{\rm 3/2} \sim 10^2$ GeV, its lifetime would lie in the
range $10^{22}-10^{24}$ sec which is well above the age of the
universe: $t_0 \simeq 3.2 \ 10^{17}$ sec.

As a first conclusion, it is interesting to note that a scenario
with trilinear RPV interactions of type $\lambda^{( \prime
)}_{ijk}$ addressing simultaneously the neutrino mass and dark
matter problems is conceivable. In the following, we will address
this double problem within a different supersymmetric scenario
where the neutrino mass is generated via bilinear and trilinear
RPV couplings of type $\mu_i$ and $\lambda_i$, respectively, which
do not induce too large decay widths of the dark matter candidate.
In contrast with the $\lambda^{( \prime )}_{ijk}$ case, the
present framework deserves a more general treatment in the sense
that the neutrinos are mixed with neutral gauginos/higgsinos so
that the neutrino constraints and the neutralino ones are
correlated. In fact, the mixing is generally so small that it does
not affect significantly the gaugino/higgsino mass matrix, but the
precise values of the several parameters entering this matrix
crucially determine the small induced neutrino masses.

Nevertheless, let us finish this part by commenting on the fact
that even if the $\lambda^{( \prime )}_{ijk}$ couplings of the order of weak Yukawa couplings
cannot induce dangerous $\tilde G$ decay channels, they could contribute
partially to the neutrino masses through one-loop diagrams. Here,
we concentrate on the dominant tree-level contributions to
the neutrino masses and hence leave for the future a precise
and complete fit of neutrino data at loop-level.

\subsection{Neutralino mass matrix}
\label{NeutMassMat}

The neutralino mass terms read, \bea \label{LAGmass} {\cal
L}^m_{\tilde \chi^0} = -\frac{1}{2} \Psi^{0^T} {\cal M}_{\tilde
\chi^0} \Psi^0 + {\rm h.c.} \eea in the basis $\Psi^{0^T}$
$\equiv$ $(\tilde B^0, \tilde W^0_3, \tilde h^0_d, \tilde h^0_u,
\tilde s, \nu_i)$, where $\tilde h^0_{u,d}$ ($\tilde s$) are the
fermionic components of the superfields $H_{u,d}^0$ ($S$) and
$\nu_i$ [$i=1,2,3$] denote the neutrinos. In Eq.~(\ref{LAGmass}),
the neutralino mass matrix is given, in a generic basis (where
$\langle \tilde \nu_i \rangle \equiv v_i \neq 0$, $\mu_i \neq 0$
and $\lambda_i \neq 0$, as will be discussed later), by
\bea \label{CHImass} {\cal M}_{\tilde
\chi^0} = \left(\begin{array}{cc}
{\cal M}_{\rm NMSSM}  & \xi_{\rpv}^{T} \\
\xi_{\rpv} & {\bf 0}_{3 \times 3}
\end{array} \right) ,
\eea

where ${\cal M}_{\rm NMSSM}$ is the neutralino mass matrix
corresponding to the NMSSM. For the latter mass matrix,
we assume $v_i \ll v_{u,d}$, so that $v^2 = v_u^2+v_d^2+
\sum^3_{i=1} v_i^2 = 2c_{\theta_W}^2m_Z^2/g^2 \simeq (175
\mbox{GeV})^2$. In the mass matrix below, $s$ and $c$ stand for
sine and cosine, respectively. \bea \label{NMSSMmass} {\cal
M}_{\rm NMSSM} = \left(\begin{array}{cccccc}
M_{1}&0 & -m_Z \ s_{\theta_W} \ c_\beta & m_Z \ s_{\theta_W} \ s_\beta & 0\\
0 & M_{2} & m_Z \ c_{\theta_W} \ c_\beta & -m_Z \ c_{\theta_W} \ s_\beta  &0\\
-m_Z \ s_{\theta_W} \ c_\beta & m_Z \ c_{\theta_W} \ c_\beta & 0 &-\mu&-\lambda v_u\\
m_Z \ s_{\theta_W} \ s_\beta &-m_Z \ c_{\theta_W} \ s_\beta&-\mu&0&-\lambda v_d+\sum^3_{i=1} \lambda_i v_i \\
0&0&-\lambda v_u &-\lambda v_d+\sum^3_{i=1} \lambda_i v_i & 2(\kappa \langle s \rangle + \mu') \\
\end{array} \right) .
\eea

Above, $M_1$ ($M_2$) is the soft supersymmetry breaking mass of
the bino (wino), $\tan \beta=v_u/v_d=\langle h^0_u \rangle
/\langle h^0_d \rangle$ (with $c_\beta = \cos \beta$ and $s_\beta = \sin \beta$), and $\mu = \lambda \langle s \rangle$.
The $\mu'$ term appears only within the GMSB framework [see Eq.(\ref{GMSB})].

We assume for simplicity that $\lambda$, $\kappa$ and the soft
supersymmetry breaking parameters are all real.

In Eq.~(\ref{CHImass}), $\xi_{\rpv}$ is the RPV part of the
matrix mixing neutrinos and neutralinos: \bea \label{RPVmass}
\xi_{\rpv} = \left(\begin{array}{ccccc}
-\frac{g'v_1}{\sqrt{2}} & \frac{gv_1}{\sqrt{2}} & 0 & \mu_1 +\lambda_1 \langle s \rangle & \lambda_1 v_u  \\
-\frac{g'v_2}{\sqrt{2}} & \frac{gv_2}{\sqrt{2}} & 0 & \mu_2 +\lambda_2 \langle s \rangle & \lambda_2 v_u  \\
-\frac{g'v_3}{\sqrt{2}} & \frac{gv_3}{\sqrt{2}} & 0 &
\mu_3+\lambda_3 \langle s \rangle & \lambda_3 v_u
\end{array} \right).
\eea

$g$ and $g'$ being the SU(2) and U(1) gauge couplings. We restrict
ourselves to the situation where $v_i/v_{u,d} \ll 1$ (as before),
$|\mu_i/\mu| \ll 1$ and $|\lambda_i/\lambda| \ll 1$ so that {\it
(i)} no considerable modifications of the NMSSM scalar potential
are induced by the additional bilinear and trilinear terms in
superpotential (\ref{Wgeneric}), {\it (ii)} the
neutralino-neutrino mixing is suppressed, leading to sufficiently
small neutrino masses; it is remarkable, as mentioned above, that the
necessary order of magnitude, $v_i/v_d \sim \mu_i/\mu \sim
\lambda_i/\lambda \sim 10^{-5}-10^{-7}$, corresponds typically to
the hierarchy between the electron mass and the top quark mass (or
equivalently the EW symmetry breaking scale).

\subsection{The various (s)lepton mixings}
\label{VariousMix}

\noindent \textbf{Chargino/neutralino-lepton mixing:} The
condition for generating two non-vanishing and non-degenerate
neutrino mass eigenvalues at tree level is to ensure
simultaneously \bea \label{condition} \mu_i\neq 0  \ \mbox{and} \
\Lambda_i\neq 0, \eea where $\Lambda_i= \langle s \rangle
(\lambda_i+\lambda \frac{v_i}{v_d})$. The respective roles of
these effective parameters is reflected e.g. in the sub-matrix
(\ref{RPVmass}) mixing the neutrinos with neutral NMSSM states.

Since the $H_d$ and $L_i$ superfields possess the same gauge
quantum numbers, one can freely rotate $L_\alpha=(H_d,L_i)$
[$\alpha=0,\dots,3$] through $SU(4)$ matrices by a redefinition of
fields (for examples of unitary matrix associated to $SU(4)$
transformations, see Ref.~\cite{Hall} or Ref.~\cite{PhysRep} in Section 2.1.4).
Motivated by the condition (\ref{condition}), we work in a
general basis where the $\mu_i L_i H_u$ terms are non-vanishing.
One could choose a specific
basis of $L_\alpha$ superfields where the $\lambda_i$ couplings
vanish, but then in general (i.e. assuming no particular
correlation between the $\lambda_i$ and $B_i$, $\tilde m^2_{d,i}$
values) some $v_i$ are generated due to the destabilization of the
scalar potential by terms linear in $\tilde \nu_i$ originating from the
RPV soft scalar bilinear terms $B_i h_u \tilde L_i$, $\tilde
m^2_{d,i} h^\dagger_d \tilde L_i$ \footnote{Similarly, there exists a
trilinear RPV soft term, arising in the NMSSM, which involves the singlet scalar field;
this one is of the form $s h_u \tilde L_i$ in the Lagrangian \cite{rpNMSSM1}.
Its direct effect on the present study will be discussed
in Section \ref{SneutStab}.} after translation of the Higgs fields: $h^0_u \to h^0_u +
v_u/\sqrt{2}$ ($h^0_d \to h^0_d + v_d/\sqrt{2}$). Reciprocally,
a field basis where the $v_i$ vanish generally leads to
non--zero $\lambda_i$ couplings. In conclusion, the condition
(\ref{condition}) is generally fulfilled.
\\ \\
\noindent \textbf{Higgs-slepton mixing:} The physical amount of
Higgs-slepton mixing is parametrized by the basis-independent
angle between the 4-vectors $v_\alpha=(v_d,v_i)$ and
$B_\alpha=(B_d,B_i)$, where $B_d$ is the soft parameter entering
the biscalar term $B_d h_u h_d$ \cite{PhysRep}. We assume this
angle sufficiently small so that the additional potentially
dangerous decay channels $\tilde G, \tilde \chi^0_1 \to H^\pm
\ell^{\mp} \ / \ h^0 \nu$ and $\tilde \nu \to f\bar{f},VV$
[$f\equiv$ fermion, $V\equiv$ Vector boson] play no role in the
present context of a {\it long-lived} LSP dark matter candidate of
type $\tilde G$, $\tilde \chi^0_1$ or $\tilde \nu$.
\\ \\
\noindent \textbf{Charged lepton mixing:} The charged leptons also
mix with the Wino (via the $v_i$'s) and the charged higgsino
(controlled similarly by the $\mu_i$ and $\lambda_i$ parameters).
However the detail of this mixing also relies on the determination
of the precise value of each Yukawa coupling constant
$Y^\ell_{ij}$ for the charged leptons [see Eq.(\ref{WNMSSM})]
which requires a complete scenario of flavor taking into account
all the experimental constraints [from neutrino oscillations and
lepton flavor violating reactions] on leptonic mixing angles
($U_{PMNS}$ matrix \cite{PMNS}), a task beyond the scope of the
present work. Therefore, we will not explicitly work out the
chargino-charged lepton mixing which is expected to be of
comparable amount as the neutralino-neutrino mixing involving
similar matrix structures and parameter orders~\cite{FluxGamma}.
Concerning the present cosmological context, this implies for
instance that the width for the decay channel $\tilde G \to W^\pm
\ell^{\mp}$ is close to the one for $\tilde G \to Z^0 \nu$, so
that our principal conclusions on the relative stability of the
gravitino LSP based on the latter channel will not be modified by
the former one \footnote{Besides, note that the contributions to
the charged lepton masses should be of the order of magnitude of
the neutrino mass scale and in turn negligible compared to the
direct Yukawa contributions.}.

\section{Sneutrino LSP}
\label{sneutrino}

\subsection{RPV sneutrino decays}

In case of a sneutrino LSP decaying through the RPV couplings of
type $\Lambda_i$ and $\mu_i$, the sneutrino can exclusively decay
into neutrinos via the two types of Feynman diagrams drawn in
Fig.(\ref{FIGSNEUT}) or into two charged leptons.
The associated decay widths are given below.

For that purpose, we need some definitions.
First, the sneutrino LSP, noted $\tilde{\nu}_1$, corresponds to the lightest of the
three sneutrino mass eigenstates, keeping in mind that arbitrary non-universal
conditions on scalar soft masses and the dependence of sneutrino mass
running e.g. on flavor-dependent $\lambda_i$ parameters lead generally
to a non-degeneracy of $m_{\tilde{\nu_i}}$ ($i=1,2,3$) eigenvalues.

We also need to introduce the matrix $N_{\alpha\beta}$ which is defined as follows.
For convenience, we change the order (w.r.t. Section \ref{NeutMassMat}) of
the fields in the weak basis [for the rest of the paper]: $\Psi^{0^T} = (\tilde B^0,\tilde W^0_3,\tilde
h^0_u,\tilde h^0_d,\tilde s,\nu_i)$ where as before the index
$i=1,2,3$, used for compact notations, corresponds
to the three flavor states: $\nu_e$, $\nu_\mu$ and
$\nu_\tau$ respectively. After dia\-gonalization, the eigenstates are ordered in
the mass basis according to $\Upsilon^{0^T} =
(\nu_j^m,\tilde{\chi}_1^0,\tilde{\chi}_2^0,\tilde{\chi}_3^0,\tilde{\chi}_4^0,\tilde{\chi}_5^0)$
where $\nu_j^m$ denotes the three neutrino mass eigenstates
$(j=1,2,3)$ and the five $\tilde{\chi}^0$'s are the NMSSM
neutralinos. Now, the unitary transformation matrix
$N_{\alpha\beta}$ acts as $\Upsilon^0_\alpha = N_{\alpha\beta}
\Psi^{0}_\beta$ with $\alpha,\beta=\{ 1 , \dots , 8 \}$.

\begin{figure}[h]
\vskip 1.0cm
\begin{center}
\includegraphics[scale=0.45]{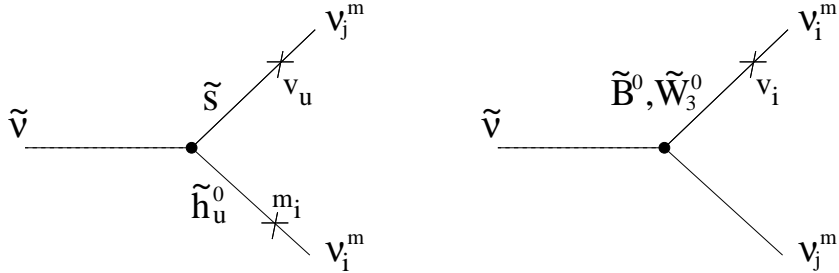}
\caption{Feynman diagrams for the sneutrino decay channel in two
neutrinos, $\tilde{\nu}\rightarrow \nu_i^m \bar{\nu}_j^m$. We use
the effective quantity $m_i = \mu_i +\lambda_i \langle s \rangle$
to parametrize a direct $\mu_i$ mixing term effect combined with a
$\langle s \rangle$ vev insertion. $v_u$ denotes the up Higgs vev
insertion and $v_i$ symbolizes the sneutrino vev insertion
($i=1,2,3$).} \label{FIGSNEUT}
\end{center}
\vspace*{5mm}
\end{figure}

\begin{itemize}

 \item  Sneutrino decay into two neutrinos $\nu_i^m \bar \nu^m_j$ (phase space factors involving neutrino over sneutrino masses are
 neglected):

\beq\label{Sneutnunu}
\Gamma(\tilde{\nu}_1\rightarrow \nu_i^m \bar{\nu}_j^m)
=\frac{m_{\tilde{\nu}_1}}{16\pi} \
\left| \sum_{k=1}^3 \frac{{\cal N}_{1k}}{1+\delta_{ij}} \left( \left[ \lambda_{k} N_{i3} N_{j5} + \frac{g}{\sqrt{2} cos\theta_W} N_{j\{k+5\}} U_{\nu_i^m\tilde{Z}} \right] +[i \leftrightarrow j]  \right) \right|^{2}
\eeq
where $i,j=\{1,2,3\}$ run over the neutrino eigenstate indexes,
${\cal N}_{1k}$ are elements of the model-dependent
sneutrino basis transformation matrix ($1$ is for the lightest sneutrino eigenstate and $k=\{1,2,3\}$ is for the sum over sneutrino flavor states),
$U_{\nu_i^m\tilde{Z}}=-N_{i1}sin\theta_{W} + N_{i2}cos\theta_{W}$
and `$+[i \leftrightarrow j]$' indicates that the same expression in brackets must be added by switching $i$ with $j$.

\item Sneutrino decay into two charged leptons $\ell_i^+ \ell^-_j$ (phase space factors involving lepton over sneutrino masses are also
negligible here):

\beq \Gamma(\tilde{\nu}_1\rightarrow \ell_i^{+}  \ell_j^{-})
=\frac{m_{\tilde{\nu}_1}}{16\pi} \ g^2 \ \left| \sum_{k=1}^3 \frac{{\cal N}_{1k}}{1+\delta_{ij}} \left( \left[ U_{j\{k+2\}} U_{\ell_i\tilde{W}} \right] +[i \leftrightarrow j] \right) \right|^{2}
 \eeq
where $U_{j\{k+2\}}$ is
the equivalent of neutralino rotation matrix $N$ but for charginos (here with $k=\{1,2,3\}$ for the sum over charged lepton flavor states
and $j=\{1,2,3\}$ associated to the final charged lepton eigenstate) and similarly $U_{\ell_i\tilde{W}}$ is the equivalent for charginos of $U_{\nu_i^m\tilde{Z}}$.

\end{itemize}

\subsection{Sneutrino stability}
\label{SneutStab}

It turns out that the strength of effective RPV couplings
$\Lambda_i/\langle s \rangle$ and $\mu_i$ necessary to generate
sufficiently large neutrino masses, i.e. typically $m_{\nu_1}
\gtrsim 10^{-1}-10^{-3}$ eV (in the approximate case of a unique
neutrino flavor), induces RPV-like mixings between neutrinos and
neutralinos which are too large from the cosmological point of
view. Indeed, those mixings translate into matrix elements $N_{i3} \sim
10^{-7}$, $N_{j5} \sim 10^{-7}$, $U_{\nu_i^m\tilde{Z}} \sim 10^{-7}$ and $N_{j\{k+5\}}  \sim 1$ with $i,j,k=1,2,3$ [for standard
neutralino masses $m_{\tilde{\chi}_1^0} = {\cal O}(10^2)$ GeV] so
that the total sneutrino RPV decay width
$\Gamma(\tilde{\nu}_1\rightarrow \nu_1^m \bar{\nu}_1^m) \sim
10^{-15}$ GeV is well above the critical value of $1.52 \
10^{-42}$ GeV [for a sneutrino mass: $m_{\tilde{\nu}_1}
\gtrsim 10$ GeV and assuming for now ${\cal N}_{1k} \sim 1$]. It means that the sneutrino lifetime is well
below the age of the universe $t_0$ and thus a sneutrino LSP does
not constitute a viable dark matter candidate. A source of width suppression may arise from the generic amount of the
matrix element ${\cal N}_{11}$ but its effect is at much of a few orders of magnitude (more suppression
would correspond either to a fine-tuning of parameters or to the weak breaking of a certain symmetry which should be described) so that
this cannot invalidate the above conclusion. Furthermore, the effect of the additional decay channels into two charged leptons and into
(pseudo)scalar Higgs bosons, $h^0_i h^0_j$, $h^0_i a^0_k$ or $a^0_k a^0_l$ [$i,j=1,2,3$, $k,l=2,3$] (via the soft term $s h_u \tilde L_i$
which is characteristic of the NMSSM with RPV interactions), can only increase the total sneutrino width
\footnote{In this part, we study the sneutrino RPV decays for completeness, even if a left--handed sneutrino LSP as a candidate for dark matter
has been already excluded by direct dark matter searches for most of the realistic sneutrino mass ranges.}.

\section{Neutralino LSP}
\label{neutralino}

\subsection{RPV neutralino decays}

\begin{figure}[h]
\vskip 1.0cm
\begin{center}
\includegraphics[scale=0.45]{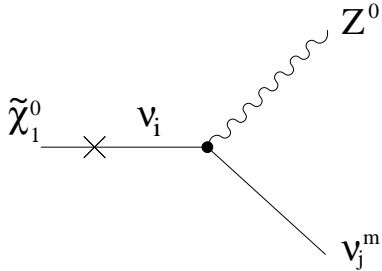}
\caption{Feynman diagram for the neutralino decay mode into the Z
boson, $\tilde \chi_{1}^{0}\rightarrow Z^{0} \nu^m_j$. The cross
allows to specify the neutralino component that is coupled (here:
a neutrino flavor state $\nu_i$).} \label{FIGNEUTRI}
\end{center}
\vspace*{5mm}
\end{figure}

\begin{figure}[h]
\vskip 1.0cm
\begin{center}
\includegraphics[scale=0.45]{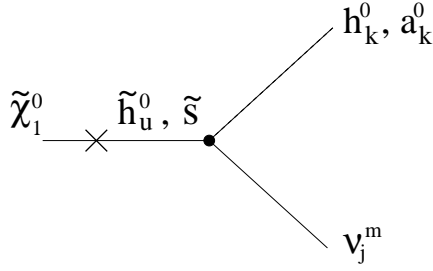}
\caption{Feynman diagram for the neutralino decay modes into
scalar Higgs fields, $\tilde \chi_{1}^{0}\rightarrow h^0_k
\nu^m_j$, and pseudoscalar Higgs fields, $\tilde
\chi_{1}^{0}\rightarrow a^0_k \nu^m_j$. The cross
indicates which neutralino components are coupled.}
\label{FIGNEUTRII}
\end{center}
\vspace*{5mm}
\end{figure}

If the LSP is the lightest neutralino and the RPV couplings of
type $\Lambda_i$ and $\mu_i$ are present, then the lightest
neutralino can only decay into the W,Z bosons or into the
(pseudo)scalar Higgs fields [as illustrated by the Feynman
diagrams of Fig.(\ref{FIGNEUTRI})-(\ref{FIGNEUTRII})] if one
restricts oneself to the dominant two-body decay channels. The obtained
partial decay widths are given as follows.


\begin{itemize}

 \item Neutralino decay into a Z boson and a neutrino $\nu^m_j$ (phase space factors involving neutrino masses are
 neglected):
\beq\label{NeutZnu} \Gamma(\tilde \chi_{1}^{0}\rightarrow Z^{0} \nu^m_j)
=\frac{1}{96\pi} \left| \sum_{i=1}^3 N_{4i} N_{ji} \right|^{2}
\frac{g^{2}}{cos^{2}\theta_{W}}m_{\tilde \chi_{1}^{0}}
\left(1-\frac{m_{Z}^{2}}{m_{\tilde \chi_{1}^{0}}^{2}} \right)^{2}
\left(1+\frac{m_{\tilde \chi_{1}^{0}}^{2}}{2m_{Z}^{2}} \right) \eeq
where $j=1,2,3$ labels the neutrino mass eigenstate and $i=6,7,8$
corresponds to the sum over neutrino flavor states.

  \item Neutralino decay into the W boson and charged lepton $\ell_j^\pm$ (the dependency on lepton masses,
which represents subleading effects, is omitted):
\beq\label{NeutWl} \Gamma(\tilde \chi_{1}^{0}\rightarrow W^{\pm}
\ell_j^{\mp}) =\frac{1}{48\pi} \left| \sum_{i=6}^8 N_{4i} U_{j\{i-3\}} \right|^{2}g^{2}m_{\tilde \chi_{1}^{0}}
\left(1-\frac{m_{W}^{2}}{m_{\tilde \chi_{1}^{0}}^{2}} \right)^{2}
\left(1+\frac{m_{\tilde \chi_{1}^{0}}^{2}}{2m_{W}^{2}} \right) \eeq
where $j=1,2,3$ labels the charged lepton mass eigenstate and $i=6,7,8$
is corresponding to the charged lepton/neutrino flavor states.

  \item Neutralino decay into scalar Higgs bosons $h^0_k$ and neutrino $\nu^m_j$:
\beq\label{Neutnuhi} \Gamma(\tilde \chi_{1}^{0}\rightarrow h^0_k
\nu^m_j) =\frac{1}{32\pi} \left| \sum_{i=6}^8 N_{ji}
\lambda_{\{i-5\}} \right|^{2} |N_{43}S_{k3}+N_{45}S_{k1}|^{2}
m_{\tilde \chi_{1}^{0}} \left(1-\frac{m_{h^0_k}^{2}}{m_{\tilde
\chi_{1}^{0}}^{2}} \right)^{2} \eeq with $j=1,2,3$ labeling the
neutrino mass eigenstate, $i=6,7,8$ corresponding to the neutrino
flavor states and $h^0_k \equiv h^0_1, h^0_2, h^0_3$ (scalar Higgs
mass eigenstates). The rotation matrix $S$ relates the real parts
of the neutral Higgs bosons and singlet scalar field to the scalar
Higgs mass eigenstates $h^0_k$ (see the precise definition in
\cite{NMHDECAY}).
We recall that the parameters $\lambda_{\{i-5\}}$ which appear
here are the NMSSM specific trilinear parameters of
Eq.(\ref{Wgeneric}).
We also mention the decays into charged Higgs fields which are
expected to be of comparable widths.

\item Neutralino decay into pseudoscalar Higgs boson $a^0_k$ and neutrino $\nu^m_j$:
\beq\label{Neutnuai} \Gamma(\tilde \chi_{1}^{0}\rightarrow a^0_k \nu^m_j)
=\frac{1}{32\pi} \left| \sum_{i=6}^8 N_{ji} \lambda_{\{i-5\}} \right|^{2} |N_{43}P_{k3}+N_{45}P_{k1}|^{2}
m_{\tilde \chi_{1}^{0}}
\left(1-\frac{m_{a^0_k}^{2}}{m_{\tilde \chi_{1}^{0}}^{2}} \right)^{2} \eeq
with $j=1,2,3$ labeling the neutrino mass eigenstate, $i=6,7,8$
corresponding to the neutrino flavor states and $a^0_k \equiv a^0_1, a^0_2$
(pseudoscalar Higgs mass eigenstates). The rotation matrix $P$ translates the imaginary parts of the neutral Higgs bosons
and singlet scalar field into the pseudoscalar Higgs mass eigenstates $a^0_k$ \cite{NMHDECAY}.

\end{itemize}

\subsection{Neutralino stability}

Based on the typical values of RPV parameters $\lambda_i$, $v_i$
and $\mu_i$ generating a neutrino mass scale $m_{\nu_1} \gtrsim
10^{-1}-10^{-3}$ eV, we find for the parameters involved in
the partial widths (\ref{NeutZnu}), (\ref{NeutWl}),
(\ref{Neutnuhi}) and (\ref{Neutnuai}): $N_{ji} \sim 1$, $N_{4i}
\sim 10^{-7}$, $U_{j\{i-3\}} \sim 1$, $\lambda_{\{i-5\}} \sim
10^{-6}$, $N_{43}S_{k3}+N_{45}S_{k1} \sim 10^{-1} - 10^{-2}$ and
$N_{43}P_{k3}+N_{45}P_{k1} \sim 10^{-1} - 10^{-2}$ with $i=6,7,8$
and $j=1,2,3$ [if neutralino masses are of the order:
$m_{\tilde{\chi}_1^0} = {\cal O}(10^2)$ GeV, and taking
$m_{\tilde{\chi}_1^0} > m_Z$]. We can see that the contributions
to the total neutralino decay width of Eqs.(\ref{Neutnuhi}) and
(\ref{Neutnuai}) are smaller than
Eqs.(\ref{NeutZnu}) and (\ref{NeutWl}).
This leads [whatever is the (pseudo)scalar Higgs spectrum] to a
total neutralino RPV decay width $\Gamma_{\rm total}(\tilde
\chi_{1}^{0}) \sim 10^{-15}$ GeV that is several orders of
magnitude above the critical value of $1.52 \ 10^{-42}$ GeV. There
are even additional contributions e.g. to the lightest neutralino decay
into $Z^{0} \nu^m_j$, of the same order (originating from the
$Z^{0} \tilde \chi_{i}^{0} \tilde \chi_{j}^{0}$ coupling), that
should slightly increase the total neutralino decay width.
Adding the decay channel into $H^\pm$, closed in most of the parts
of the parameter space, can only increase again the total
neutralino width. The conclusion is thus as for the sneutrino
case: the neutralino lifetime is much smaller than $t_0$ and hence
it does not represent a possible dark matter LSP candidate.

\begin{figure}[h]
\vskip 1.0cm
\begin{center}
\includegraphics[scale=0.4]{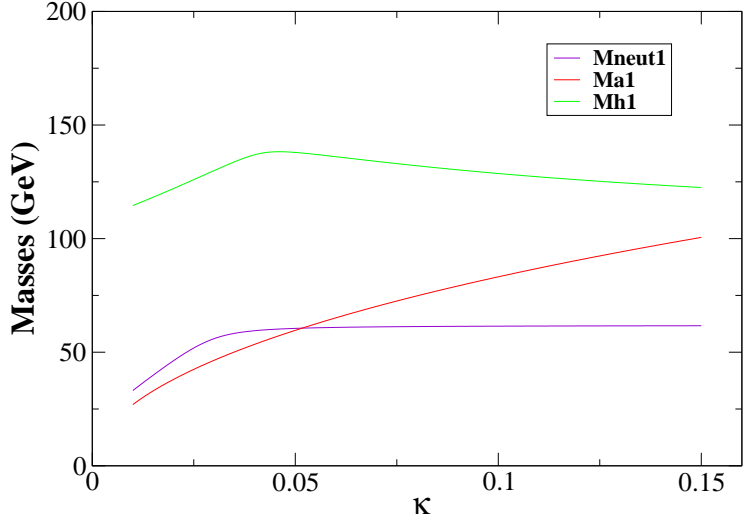}
\caption{Masses $m_{\tilde \chi_{1}^{0}}$ [purple curve],
$m_{h^0_1}$ [green curve] and $m_{a^0_1}$ [red curve] as functions
of $\kappa$. We fix the other NMSSM parameters:
tan$\beta$=2, $\lambda$=0.7,
$\mu=\lambda \langle s \rangle$=530 GeV, $M_1$=66 GeV,
$M_2$=133 GeV, $M_3$=500 GeV, $A_{\lambda}$=1280 GeV,
$A_{\kappa}$=0, $A_{t}=A_{b}=A_{\tau}$=-2.5 TeV,
$m_{\tilde{\ell}^\pm}$=200 GeV and $m_{\tilde{q}}$=1 TeV (universally). The $A$
parameters are the trilinear scalar soft supersymmetry breaking
parameters which do not affect the neutralino mass matrix. The low
values of $M_i$'s allow to get a neutralino LSP. Note also the
particularly low values of $m_{a^0_1}$, characteristic of the
NMSSM. For completeness, we give the employed RPV parameters:
$\Lambda_1/\langle s \rangle=9 \ 10^{-7}$ and $\mu_{1}=6 \
10^{-6}$ GeV.} \label{fig.neutralino1}
\end{center}
\vspace*{5mm}
\end{figure}

Nevertheless, a possibility to insure the stability of the
lightest neutralino is to restrict to domains of parameter space
where the 4 possible decay channels of Eq.(\ref{NeutZnu}), Eq.(\ref{NeutWl}),
Eq.(\ref{Neutnuhi}) and Eq.(\ref{Neutnuai}), as well as the channels into charged Higgs bosons,
are all kinematically closed [even opening only one channel is
sufficient to render the $\tilde \chi_{1}^{0}$ unstable]. Although
in the NMSSM the lightest pseudoscar Higgs boson noted $a_1$ can
be much lighter than in the MSSM, we can find realistic regions of
the NMSSM parameter space where the lightest neutralino can be
simultaneously lighter than the W, Z bosons, the lightest scalar
$h^0_1$ and pseudoscalar $a^0_1$, as well as the charged Higgs
bosons $H^\pm$, thus closing the corresponding RPV channels.
By the term `realistic', we mean that the NMSSM parameters pass
the theoretical and phenomenological constraints implemented in
the NMHDECAY program \cite{NMHDECAY} like: {\it (i)} the physical
minimum of the scalar potential is deeper than the local
unphysical minima with $\langle h^0_{u,d} \rangle = 0$ and/or
$\langle s \rangle = 0$ {\it (ii)} the running couplings
$\lambda$, $\kappa$, $Y^b$, $Y^t$ do not encounter a Landau
singularity {\it (iii)} the experimental constraints from LEP in
the neutralino, chargino and Higgs sectors are effectively
satisfied.
\\
For instance, using NMHDECAY, we have plotted the masses
$m_{\tilde \chi_{1}^{0}}$, $m_{h^0_1}$ and $m_{a^0_1}$ as a
function of the $\kappa$ coupling constant in
Fig.(\ref{fig.neutralino1}), for fixed values of the other
NMSSM parameters. We see on this figure that for $\kappa \gtrsim
0.05$, the lightest neutralino LSP is lighter than the W, Z
bosons, the lightest scalar $h^0_1$ and pseudoscalar $a^0_1$. So
the scenario of a completely stable $\tilde \chi_{1}^{0}$ LSP can
be a priori realized. Note that in such a scenario the RPV couplings can now be chosen
freely to satisfy the neutrino constraints, since those couplings
have a negligible impact on the neutralino mass spectrum.

We remark that there exist various triangular one-loop processes, exchanging charged leptons/sleptons,
quarks/squarks or charged leptons/W bosons (through bilinear or trilinear RPV couplings
\footnote{The contributions from trilinear RPV interactions were estimated in Ref.~\cite{Lola}.}), which contribute
all to the decay channel $\tilde \chi_{1}^{0} \rightarrow \gamma \nu^m_i$. Such a channel is always
kinematically open in the NMSSM, even in the case of the specific spectrum discussed just above, and
could thus render the neutralino unstable from the cosmological scale point of view. Assuming that no
important destructive interferences occur among the several triangular contributions mentioned above,
rough estimates indicate that the suppression provided by RPV terms [reproducing the correct
neutrino mass scale] and by loop factors are not sufficient
to make the LSP neutralino stable with respect to $t_0$ (whatever is the neutralino composition). However,
precise loop calculations of all contributing reactions, beyond the scope of our study,
are necessary to conclude definitively on this aspect, taking into account notably effects of heavy
sfermion loop factors.
\\ The other important final comment is that three-body decay channels like $\tilde \chi_{1}^{0} \rightarrow \ell^+ \ell^- \nu^m$
(via an off-shell Z/W boson \cite{ValleI,ValleII} or an off-shell Higgs boson \cite{ValleI}) are systematically open. From order of 
magnitude estimates, it tuns out that such channels are expected to render the lightest neutralino clearly unstable.

\section{Gravitino LSP}
\label{gravitino}

\subsection{RPV gravitino decays}

\begin{figure}[h]
\vskip 1.0cm
\begin{center}
\includegraphics[scale=0.45]{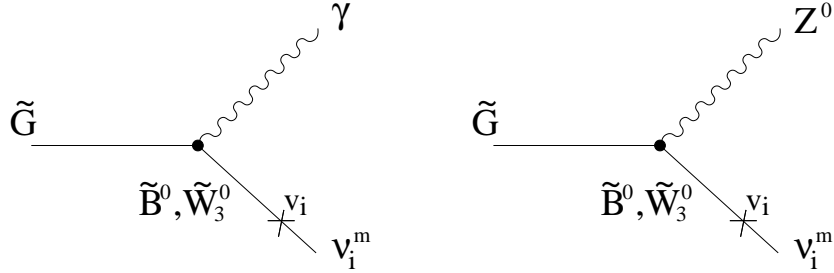}
\caption{Feynman diagrams for the gravitino decay processes into a
photon, $\tilde{G}\rightarrow \gamma \nu_i^m$, and a Z boson,
$\tilde{G}\rightarrow Z^0 \nu_i^m$. $v_i$ denotes a sneutrino
vev insertion.} \label{FIGGRAVI}
\end{center}
\vspace*{5mm}
\end{figure}

\begin{figure}[h]
\vskip 1.0cm
\begin{center}
\includegraphics[scale=0.45]{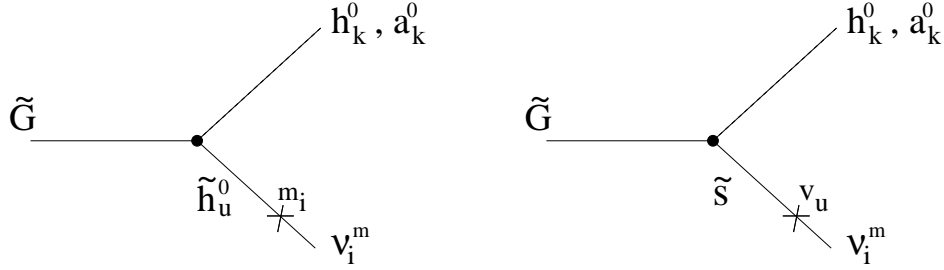}
\caption{Feynman diagrams for the gravitino decay processes into
scalar Higgs fields, $\tilde{G}\rightarrow h^0_k \nu_i^m$, and
pseudoscalar Higgs fields, $\tilde{G}\rightarrow a^0_k \nu_i^m$.
We use the symbolic notation $m_i = \mu_i +\lambda_i \langle s \rangle$
combining a direct $\mu_i$ mixing and a
$\langle s \rangle$ insertion. $v_u$ indicates an up Higgs vev
insertion.} \label{FIGGRAVII}
\end{center}
\vspace*{5mm}
\end{figure}

In the case of a gravitino LSP and RPV couplings of type
$\Lambda_i$ and $\mu_i$, the gravitino can decay into
the EW gauge bosons and the (pseudo)scalar Higgs fields [as shown
in Fig.(\ref{FIGGRAVI})-(\ref{FIGGRAVII})]. We give below the
obtained associated partial decay widths.
\\ All formulas given for the gravitino width, and above for sneutrino and neutralino widths,
result from original calculations. Nevertheless, some of the gravitino decay
amplitudes, into $\gamma$, $Z^0$, $W^+$ and $h^0_k$, were computed with
other conventions
by the authors of Ref.~\cite{Buch,Flux,Grefe} in the case of the MSSM
(see also Ref.~\cite{Moroi} for a more generic calculation approach). Note in
particular that within the
present framework of the NMSSM, the numerical results for the gravitino decay
into the pseudoscalar $a^0_1$ -- which can be quite light there -- might differ significantly
from the MSSM and thus decrease dangerously the gravitino lifetime.
Moreover, within the NMSSM, new contributions arise for the decays
$\tilde{G}\rightarrow h^0_k \nu_i^m$ and $\tilde{G}\rightarrow a^0_k \nu_i^m$
due to the presence of the $s$ superfield [see the right Feynman diagram of Fig.(\ref{FIGGRAVII})].
Finally, in the NMSSM, the neutralino mass matrix (and in turn the matrix elements $N_{ij}$
involved in gravitino amplitudes) takes a specific form, especially in the context of the GMSB
(see Section \ref{NeutMassMat}).


\begin{itemize}

\item Gravitino decay into the photon $\gamma$ and neutrino $\nu_i^m$
-- the dependency on the neutrino mass is omitted (see first calculation in Ref.~\cite{Takayama}):
\beq\label{Gravgammanu} \Gamma(\tilde{G}\rightarrow \gamma
\nu_i^m) =\frac{1}{64\pi}|U_{\nu_i^m\tilde{\gamma}}|^{2}
\frac{m_{\rm 3/2}^{3}}{M_{Pl}^{2}} \eeq where
$U_{\nu_i^m\tilde{\gamma}}=N_{i1}cos\theta_{W} +
N_{i2}sin\theta_{W}$, $i=1,2,3$ labels the neutrino mass
eigenstate and $1$ ($2$) corresponds to the $\tilde B^0$ ($\tilde
W^0_3$) component. $M_{Pl}$ is the reduced Planck mass: $M_{Pl}
\simeq 2.4 \ 10^{18}$ GeV.

\item Gravitino decay into the Z boson and neutrino $\nu_i^m$
(see original calculation in Ref.~\cite{FluxGamma}):
\beq\label{GravZnu} \Gamma(\tilde{G}\rightarrow Z^0 \nu_i^m)
=\frac{1}{64\pi}|U_{\nu_i^m\tilde{Z}}|^{2}
\frac{m_{\rm 3/2}^{3}}{M_{Pl}^{2}}
\left(1-\frac{m_{Z}^{2}}{m_{\rm 3/2}^{2}} \right)^{2} \left(1 +
\frac{2}{3}\frac{m_{Z}^{2}}{m_{\rm 3/2}^{2}} +
\frac{1}{3}\frac{m_{Z}^{4}}{m_{\rm 3/2}^{4}} \right) \eeq where
$i=1,2,3$ labels the neutrino mass eigenstate and $1$ ($2$)
corresponds to the $\tilde B^0$ ($\tilde W^0_3$) component.

\item Gravitino decay into the W boson and charged lepton $\ell^{\mp}_i$
(see original calculation in Ref.~\cite{FluxGamma}):
\beq\label{GravWl} \Gamma(\tilde{G}\rightarrow W^{+} \ell^{-}_i)
=\frac{1}{32\pi}|U_{\ell_i\tilde{W}}|^{2}
\frac{m_{\rm 3/2}^{3}}{M_{Pl}^{2}}
\left(1-\frac{m_{W}^{2}}{m_{\rm 3/2}^{2}} \right)^{2} \left(1 +
\frac{2}{3}\frac{m_{W}^{2}}{m_{\rm 3/2}^{2}} +
\frac{1}{3}\frac{m_{W}^{4}}{m_{\rm 3/2}^{4}} \right) \eeq where
numerically $U_{\ell_i\tilde{W}}$ is expected to be of a
comparable order to $U_{\nu_i^m\tilde{Z}}$ as already discussed.
Similarly, here, $i=1,2,3$ labels the charged lepton mass
eigenstate.

  \item Gravitino decay into the scalar Higgs boson $h^0_k$ and neutrino $\nu_i^m$
  \beq\label{Gravhinu}
\Gamma(\tilde{G}\rightarrow h^0_k \nu_i^m)
=\frac{1}{384\pi}|N_{i3}S_{k1}+N_{i4}S_{k2}+N_{i5}S_{k3}|^{2}
\frac{m_{\rm 3/2}^{3}}{M_{Pl}^{2}}
\left(1-\frac{m_{h^0_k}^{2}}{m_{\rm 3/2}^{2}} \right)^{4} \eeq with
$i=1,2,3$ labeling the neutrino mass eigenstate, the numbers $3$,
$4$, $5$ corresponding respectively to the $\tilde h^0_u$, $\tilde
h^0_d$, $\tilde s$ components and $1$, $2$, $3$ respectively to
the real $h^0_u$, $h^0_d$, $s$ components.

  \item Gravitino decay into the pseudoscalar Higgs boson $a^0_k$ and neutrino $\nu_i^m$
 \beq\label{Gravainu}
\Gamma(\tilde{G}\rightarrow a^0_k \nu_i^m)
=\frac{1}{384\pi}|N_{i3}P_{k1}+N_{i4}P_{k2}+N_{i5}P_{k3}|^{2}
\frac{m_{\rm 3/2}^{3}}{M_{Pl}^{2}}
\left(1-\frac{m_{a^0_k}^{2}}{m_{\rm 3/2}^{2}} \right)^{4} \eeq still
with $k \equiv 1,2$. Similarly, $i=1,2,3$ labels the neutrino mass
eigenstate, the numbers $3$, $4$, $5$ correspond respectively to
the $\tilde h^0_u$, $\tilde h^0_d$, $\tilde s$ components and $1$,
$2$, $3$ respectively to the imaginary $h^0_u$, $h^0_d$, $s$ components.

 \item Gravitino decay into the charged Higgs boson $H^{\pm}$ and charged lepton $\ell_i^{\mp}$
 \beq\label{GravHell}
\Gamma(\tilde{G}\rightarrow H^{+} \ell_i^{-})
=\frac{1}{384\pi} | U_{i2} |^2 \frac{m_{\rm 3/2}^{3}}{M_{Pl}^{2}}
\left(1-\frac{m_{H^{+}}^{2}}{m_{\rm 3/2}^{2}} \right)^{4} \eeq
$i=1,2,3$ labels the charged lepton mass eigenstate.
$U_{i2}$ represents the components
of the three charged lepton eigenstates $\ell_i^\pm$ into the charged higgsino
$\tilde{h}^{\pm}$.

\end{itemize}

We note that the partial widths for the charge conjugated final
states are equal, which means for instance that
$\Gamma(\tilde{G}\rightarrow \gamma \bar{\nu}) =
\Gamma(\tilde{G}\rightarrow \gamma \nu)$. Later, we will thus
refer to the total gravitino decay width as: $\Gamma_{\rm
total}(\tilde{G}) = 2 [
\Gamma(\tilde{G}\rightarrow \gamma \nu^m) +
\Gamma(\tilde{G}\rightarrow Z^0 \nu^m) +
\Gamma(\tilde{G}\rightarrow W^{+} \ell^{-}) +
\Gamma(\tilde{G}\rightarrow h^0 \nu^m) +
\Gamma(\tilde{G}\rightarrow a^0 \nu^m) +
\Gamma(\tilde{G}\rightarrow H^{+} \ell^{-})
] $.

\subsection{Gravitino stability}

\subsubsection{Scenario I: almost all decay channels open}
\label{scenarioI}

Let us first consider a NMSSM scenario which is {\it a priori}
dangerous from the cosmological point of view: we choose an heavy
gravitino, so that the 5 first of the 6 types of decay channels described above
are kinematically open (parameters allowing only the decays into
the first $h^0_1$ and $a^0_1$), and a quite light pseudoscalar
$a^0_1$ (as is possible in the NMSSM), tending to increase the
phase space for the partial width $\Gamma(\tilde{G}\rightarrow
a^0_1 \nu_1^m)$.

\begin{figure}[h]
\vskip 1.0cm
\begin{center}
\epsfig{figure=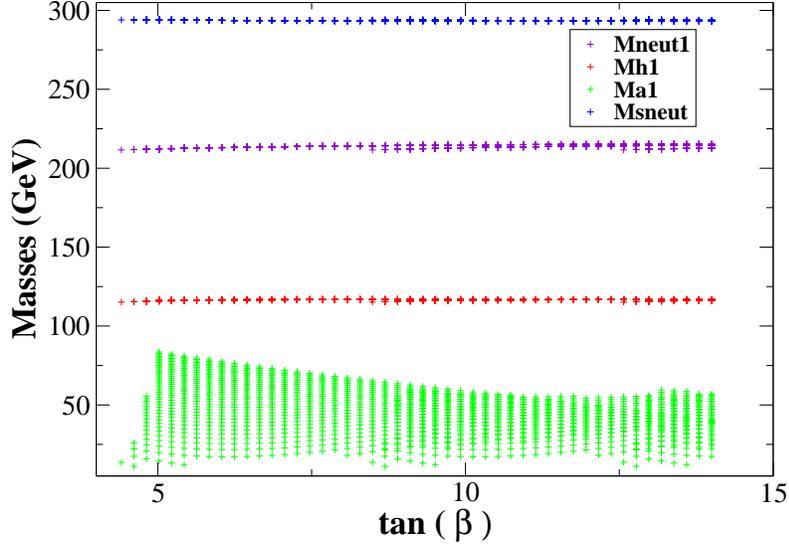,width=9cm,angle=270}
\caption{Masses $m_{\tilde \chi_{1}^{0}}$ [purple points],
$m_{\tilde\nu}$ (universal) [blue points], $m_{h^0_1}$ [red points] and
$m_{a^0_1}$ [green points] as functions of tan$\beta$. We use:
$\lambda$=0.3, $\kappa$=0.35, $\mu$=237 GeV, $M_1$=300
GeV, $M_2$=600 GeV, $M_3$=2 TeV, $A_{\kappa}$=-30 GeV,
$A_{t}=A_{b}=A_{\tau}$=-2500 GeV, $m_{\tilde{\ell}^\pm}$=300 GeV (universal),
$m_{\tilde{q}}$=1 TeV. The points are obtained from a scan
performed on the two parameters tan$\beta = 4 \rightarrow 14$ and
$M_A = 250 \rightarrow 300$ GeV, using the NMHDECAY code
\cite{NMHDECAY}. The chosen RPV parameters are: $\Lambda_1/\langle
s \rangle=2 \ 10^{-6}$ and $\mu_{1}=10^{-5}$ GeV.}
\label{fig.gravitino1}
\end{center}
\vspace*{5mm}
\end{figure}

First, the considered region of the parameter
space must be such that the gravitino is the LSP. We thus take
rather large values for the gaugino masses $M_1$, $M_2$ and $M_3$:
$M_1$=300 GeV, $M_2$=600 GeV, $M_3$=2 TeV to push the neutralino
mass to higher values. For the same reason, we take
$m_{\tilde{\ell}^\pm}$=300 GeV and $m_{\tilde{q}}$=1 TeV. We fix
the other NMSSM parameters at: $\lambda$=0.3, $\mu$=237
GeV, $\kappa$=0.35, $A_{\kappa}$=-30 GeV,
$A_{t}=A_{b}=A_{\tau}$=-2500 GeV (trilinear soft parameters). We allow tan$\beta$ and $M_A$ to
vary accordingly to a scan performed for tan$\beta = 4 \rightarrow
14$, $M_A = 250 \rightarrow 300$ GeV. $M_A$ represents the
pseudoscalar mass in the MSSM but it is an effective parameter in
the NMSSM. It is somewhat equivalent to the second (pseudo)scalar
masses in the NMSSM and to the charged Higgs mass which are almost
degenerated in this model: $M_A \simeq m_{h^0_2}, m_{a^0_2},
m_{H^{\pm}}$. It is related to the other ones via the minimization
equations. In the Fortran code NMHDECAY, we can choose between $M_A$
or $A_{\lambda}$ as input parameters.
This scan is performed through the NMHDECAY code so that the
generic NMSSM constraints mentioned above are satisfied for the
selected parameters.
\\ As a first step, we present in Fig.(\ref{fig.gravitino1}) the
output masses $m_{\tilde \chi_{1}^{0}}$, $m_{\tilde\nu_1}$,
$m_{h^0_1}$ and $m_{a^0_1}$, obtained via this scan, as functions
of tan$\beta$.

Among the possible points of Fig.(\ref{fig.gravitino1}), we choose
the one corresponding to tan$\beta$=8 and $M_A$=275 GeV leading to
$m_{\tilde \chi_1^0}$= 219.5 GeV, $m_{\tilde{\nu}}$=293.5 GeV,
$m_{h^0_1}$= 116.7 GeV, $m_{h^0_2}$= 259.9 GeV, $m_{h^0_3}$= 548.8
GeV, $m_{a^0_1}$= 25.3 GeV, $m_{a^0_2}$= 303.5 GeV and
$m_{H^{\pm}}$= 265.5 GeV. The gravitino mass is fixed at $m_{\rm 3/2}$=200 GeV so that $\tilde G$ is well the LSP. Then the
channels $\tilde{G}\rightarrow h^0_k \nu^m$,
$\tilde{G}\rightarrow a^0_k \nu^m$ [with $k \geq 2$] and
$\Gamma(\tilde{G}\rightarrow H^{\pm} \ell^{\mp})$ are
kinematically closed but all the other ones are open.

Now, in the one lepton flavor approximation we choose the RPV parameter
values $\Lambda_1/\langle s \rangle=2 \ 10^{-6}$ and
$\mu_{1}=10^{-5}$ GeV leading to $m_{\nu_1}^2 =1.04 \ 10^{-22}$
GeV$^2$ which is reasonable from the point of view of
experimental neutrino data.

Finally, for the chosen NMSSM parameters and RPV couplings, the
induced RPV neutrino-neutralino mixings give rise to the partial
widths $\Gamma(\tilde{G}\rightarrow \gamma \nu^m) = 1.95 \
10^{-47}$ GeV, $\Gamma(\tilde{G}\rightarrow Z^0 \nu^m) = 8.80 \
10^{-47}$ GeV, $\Gamma(\tilde{G}\rightarrow W^{+} \ell^{-}) = 1.94
\ 10^{-46}$ GeV, $\Gamma(\tilde{G}\rightarrow h_1^0 \nu^m) = 9.47
\ 10^{-45}$ GeV and $\Gamma(\tilde{G}\rightarrow a_1^0 \nu^m) =
2.08 \ 10^{-46}$ GeV. It is important to note that the relative
smallness of $\Gamma(\tilde{G}\rightarrow \gamma \nu^m)$ reflects
in particular the smallness of the photino component for the
neutrino mass eigenstate. The corresponding total gravitino width
is $\Gamma_{\rm total}(\tilde{G}) = 2.00 \ 10^{-44}$ GeV giving
rise to a gravitino lifetime $\tau_{\tilde{G}} \simeq \ 76.3 \
t_0$. This result illustrates the feature that within the NMSSM,
and for RPV couplings that generate the correct neutrino mass
scale, a gravitino LSP is stable (with respect to $t_0$) even for
a gravitino mass up to $200$ GeV typically in agreement with
supergravity (or mixed with GMSB) scenarios.
\\
Indeed, more generally speaking $\Gamma_{\rm total}(\tilde{G})$
remains of the same order of magnitude as for the above NMSSM
parameters, if $m_\nu \sim 10^{-11}$ GeV (as imposed by data) and
$m_{\tilde \chi_1^0} \sim m_{\rm 3/2} \sim 10^2$ GeV (as natural
in supergravity-like models). The reason, taking e.g.
$\Gamma(\tilde{G}\rightarrow \gamma \nu^m)$, is that the orders of
$m_\nu$ and $m_{\tilde \chi_1^0}$ fix the order of
$U_{\nu^m\tilde{\gamma}}$ (see Eq.(6) in Ref.~\cite{Takayama}) and
in turn the partial gravitino decay width ({\it c.f.}
Eq.(\ref{Gravgammanu})) once $m_{\rm 3/2}$ is chosen. Furthermore
one expects $U_{\nu^m\tilde{Z}} \simeq U_{\ell \tilde{W}} \simeq
U_{\nu^m\tilde{\gamma}}$ in Eq.(\ref{GravZnu})-(\ref{GravWl})
\cite{FluxGamma}.
Similarly, the partial decay widths into the (lightest) neutral
Higgs bosons are fixed by $m_{\rm h^0_1}$, $m_{\rm a^0_1}$,
$m_{\rm 3/2}$ as well as the amount of neutrino components in the
higgsino and singlino ({\it c.f.} Fig.(\ref{FIGGRAVII})) [and thus
by $m_\nu$ and $m_{\tilde \chi_1^0}$]. If kinematically open, the
decays into the $h^0_2/a^0_2$ Higgs bosons have rates of similar
order (for identical parameters). For a $a^0_1$ boson heavier than
in the above set of parameters, $\Gamma(\tilde{G}\rightarrow a_1^0
\nu^m)$ decreases but the total width remains around $\sim
10^{-44}$ GeV. Finally, if allowed, the channel
$\tilde{G}\rightarrow H^{+} \ell^{-}$ is expected to reach a width
of the same order as $\tilde{G}\rightarrow h_1^0 \nu^m$ and hence
would not significantly modify $\Gamma_{\rm total}(\tilde{G})$.
\\
It has been recently \cite{recentCHOI} pointed out that the widths
for $\tilde{G} \rightarrow \ell^{\pm} W^{\mp \star} \rightarrow
\ell^{\pm} f \bar f'$ and $\tilde{G} \rightarrow \nu^m Z^{\star}
\rightarrow \nu^m f \bar f$ together can reach $10^2 \times
\Gamma(\tilde{G}\rightarrow \gamma \nu^m)$ restricting oneself
around parameter domains where $m_{\tilde \chi_1^0} \sim m_{\rm
3/2} \sim 10^2$ GeV \footnote{This ratio result is not expected to
be significantly changed when extending the MSSM to the NMSSM.}.
It means that the new three-body decay widths can reach $\sim
10^{-45}$ GeV, from our present results, and this does not modify
$\Gamma_{\rm total}(\tilde{G})$ significantly.

\subsubsection{Scenario II: a few decay channels open}
\label{single}

To be more general than in the approach of Section
\ref{scenarioI}, we consider here the case where the gravitino is
the LSP and is lighter than the W boson as well as the
(pseudo)scalar Higgs fields [rather than considering some given
points of the parameter space]. Then the RPV channel through the
process $\tilde{G}\rightarrow \gamma \nu_i^m$ remains open.
Imposing the width $\Gamma(\tilde{G}\rightarrow \gamma \nu_i^m)$
to be smaller than the critical value of $1.52 \ 10^{-42}$ GeV
leads to the upper bound on the gravitino mass $m_{\rm 3/2}
\lesssim 3$ TeV for $\vert U_{\tilde{\gamma}\nu} \vert \sim
10^{-7}$ (see Eq.(\ref{Gravgammanu})) as imposed by the neutrino
mass scale if $m_{\tilde \chi_1^0} \lesssim 10^2$ GeV. For
example, let us consider a given point of parameter space for
which such a neutralino mass is realized. We take randomly the
point of Fig.(\ref{fig.neutralino1}) with $\kappa$=0.1. Then,
$m_{\tilde \chi_1^0}$= 67.8 GeV, $m_{\tilde{\nu}}$=193.9 GeV
(universal), $m_{h^0_1}$= 128.7 GeV and $m_{a^0_1}$= 83.2 GeV
[insuring also that the gravitino decays into the (pseudo)scalar
Higgs fields are forbidden]. For this point, one gets
$U_{\nu_i^m\tilde{\gamma}} = 1.94 \ 10^{-7}$ with
$\Lambda_1/\langle s \rangle=9 \ 10^{-7}$ and $\mu_{1}=6 \
10^{-6}$ GeV which reproduce the correct neutrino mass scale
($m_{\nu_1}^2 =5.73 \ 10^{-22}$ GeV$^2$). The exact bound
resulting from the gravitino lifetime in this precise case is
$m_{\rm 3/2} < 2.7$ TeV. Taking into account the other allowed
channels, $\tilde{G} \rightarrow \ell^{\pm} f \bar f'$ and
$\tilde{G} \rightarrow \nu^m f \bar f$ [$f\equiv$ fermions] (see Ref.~\cite{recentCHOI}), this
bound changes to $m_{\rm 3/2} \lesssim$ 650 GeV -- 3 TeV as long
as $m_{\tilde \chi_1^0} \sim 10^2$ GeV. Since we are in a
situation where $m_{\rm 3/2} < m_W$, this bound is respected. From
the general point of view, we conclude that the NMSSM with
phenomenologically interesting RPV couplings implies a gravitino
LSP which is systematically stable (with respect to $t_0$) if one
takes its mass to be smaller than $m_{W}$, $m_{h^0_1}$ and
$m_{a^0_1}$.
\\ This quite general conclusion applies to supergravity (or mixed with GMSB) scenarios where
$m_{\rm 3/2} \sim 50$ GeV is realistic, as well as to pure GMSB
scenarios in which possibly $m_{\rm 3/2} \sim 1$ eV. Indeed,
within the GMSB framework, the new form of the neutralino mass
matrix (see Section \ref{NeutMassMat}) leads to values of
$U_{\nu_i^m\tilde{\gamma}}$ extremely close to the above case [and
hence to same bounds on $m_{\rm 3/2}$] for the realistic neutrino
mass scale, if one chooses e.g. the typical characteristic value
$\mu' = 100$ GeV. Note that in GMSB, the gravitino is clearly
automatically lighter than $W$, $h^0_1$ and $a^0_1$. Note also
that the channel $\tilde{G}\rightarrow \gamma \nu_1^m$ is
systematically open (in the realistic 3 flavor case) since within
our neutrino scenario $m_{\nu_1}=0$ at the tree level.

\subsubsection{Scenario II extended to 3 neutrino flavors}
\label{subTHREE}

There exist sets of NMSSM parameters [passing the NMHDECAY
constraints] and RPV couplings which reproduce the squared
neutrino mass eigenvalue differences measured in oscillation
experiments. More precisely, the updated three-flavor analyzes
based on a global fit including results from solar, atmospheric,
reactor and accelerator oscillation experiments lead to (at the $3
\sigma$ level): $7.1 \leq \Delta m_{21}^2 \leq 8.3 \ [10^{-5}
\mbox{eV}^2]$ and $2.0 \leq \Delta m_{31}^2 \leq 2.8 \ [10^{-3}
\mbox{eV}^2]$ \cite{Valle}. In our notations e.g. $\Delta m_{31}^2
= \vert m_{\nu_3}^2 - m_{\nu_1}^2 \vert$. Values lying in those
two intervals arise for instance for the point of
Fig.(\ref{fig.neutralino1}) if $\kappa$=0.1 and \footnote{Let us
note that we have chosen the specific set (\ref{param3F}) of RPV
parameters to illustrate on an explicit example that the three
flavor neutrino data can be reproduced. Nevertheless, the orders
of magnitude in Eq.(\ref{param3F}) of the six effective RPV
parameters are general in the sense that those are imposed
systematically by the experimental ranges for $\Delta m_{21}^2$
and $\Delta m_{31}^2$ as soon as $m_{\tilde \chi_1^0} \sim 10^2$
GeV (as occurs in supergravity and GMSB models). Hence the
conclusions at the end of this Section \ref{subTHREE}, which are
based on these orders of magnitude, can also be generalized to any
region of the parameter space where
$m_{\tilde \chi_1^0} \sim 10^2$ GeV.}
\begin{eqnarray}
\mu_1=3 \  10^{-5} ~\mbox{GeV}, \ \mu_2=1.5 \  10^{-4}
~\mbox{GeV},
\ \mu_3=2.5 \  10^{-4} ~\mbox{GeV} \nonumber \\
\Lambda_1 / \langle s \rangle =8.5 \  10^{-7}, \ \Lambda_2 /
\langle s \rangle =1 \  10^{-7}, \ \Lambda_3 / \langle s
\rangle =1.5 \  10^{-7}. \label{param3F}
\end{eqnarray}
Indeed, this complete set of parameters yields the following three
neutrino mass eigenvalues at tree level:
\begin{equation}
m_{\nu_1} = 0 ~\mbox{eV}, \ m_{\nu_2} = 0.00867 ~\mbox{eV}, \
m_{\nu_3} = 0.04670 ~\mbox{eV} \label{Realistic}
\end{equation}
in agreement with the above experimental intervals for $\Delta
m_{21}^2$ and $\Delta m_{31}^2$ since in the present RPV model one
has $\Delta m_{21}^2 = m_{\nu_2}^2 = 7.51 \  10^{-5}$ eV$^2$
and $\Delta m_{31}^2 = m_{\nu_3}^2 = 2.18 \  10^{-3}$ eV$^2$.
\\ From the cosmological point of view, the neutrino mass eigenvalues in
(\ref{Realistic}) satisfy the bound extracted from WMAP and 2dFGRS galaxy
survey (depending on cosmological priors): $\sum_{i=1}^3 m_{\nu_i} \lesssim
~0.7~\mbox{eV}$ \cite{cosmobound}.
\\ Finally, the above neutrino mass eigenvalues are perfectly
compatible with the limits extracted from the tritium beta decay experiments
($95 \% \ {\rm C.L.}$): $m_\beta \leq 2.2 \ \mbox{eV} \ \mbox{[Mainz]}$ and
$m_\beta \leq 2.5 \ \mbox{eV} \ \mbox{[Troitsk]}$, this
effective mass being defined as $m^2_\beta = \sum_{i=1}^3 |U_{ei}|^2
m_{\nu_i}^2$ where $U_{ei}$ is the leptonic mixing matrix \cite{absmass}.

The point of parameter space considered in this section is the
same as the one considered in the previous section (i.e. Section
\ref{single}), namely: the point of Fig.(\ref{fig.neutralino1})
with $\kappa$=0.1, and we still assume a situation where the
gravitino LSP has a mass smaller than $m_{W}$, $m_{h^0_1}$ and
$m_{a^0_1}$ (scenario II). For this point, and with the RPV
couplings of Eq.(\ref{param3F}), we obtain for the
gravitino decay channel into a photon:
$U_{\nu_1^m\tilde{\gamma}}=-4.74 \  10^{-11}$,
$U_{\nu_2^m\tilde{\gamma}}=-6.61 \  10^{-8}$ and
$U_{\nu_3^m\tilde{\gamma}}=2.30 \  10^{-7}$,
recalling the definition
$U_{\nu_i^m\tilde{\gamma}}=N_{i1}cos\theta_{W} +
N_{i2}sin\theta_{W}$ where $i=1,2,3$ labels the neutrino mass
eigenstate and $1$ ($2$) corresponds to its $\tilde B^0$ ($\tilde
W^0_3$) component. Those numbers imply the limit $m_{\rm 3/2} <
2.5$ TeV resulting from the condition
$$
2 \ \sum_i \ \Gamma(\tilde{G}\rightarrow \gamma \bar \nu_i^m) \ < \ 1.52 \ 10^{-42} \ \mbox{GeV}.
$$
The resulting bound $m_{\rm 3/2} < 2.5$ TeV obtained here is close to the bound $m_{\rm 3/2} < 2.7$ TeV of previous
section, for the channel $\tilde{G}\rightarrow \gamma \nu^m$, which means that moving to the three flavor case should not
affect significantly the conclusions. Therefore we conclude that,
in the case of three neutrino flavors as well, a gravitino LSP is
always sufficiently stable (with respect to $t_0$) if its mass is weaker than $m_{W}$, $m_{h^0_1}$ and
$m_{a^0_1}$.

This conclusion on the 3 neutrino flavor case is not trivial in
the sense that, starting from the 1 flavor situation, the
variation of the $U_{\nu_1^m\tilde{\gamma}}$ value when extending
to 3 neutrino flavors cannot be easily predicted due to the rich
structure of the whole RPV neutralino mass matrix. Moreover, there
is no simple argument to deduce the values of
$U_{\nu_2^m\tilde{\gamma}}$ and $U_{\nu_3^m\tilde{\gamma}}$ from
the $U_{\nu_1^m\tilde{\gamma}}$ element (encoding the
neutrino-gaugino mixing) because of the multiple mixing types
between the different flavors of neutrinos and the various
neutralino states.
\\
Nevertheless, we have checked that, as expected, one obtains
numerically the following hierarchy among the following matrix elements:
$\vert N_{1j} \vert < \vert N_{2j} \vert < \vert N_{3j} \vert$
[$j=1,2$]. This is interpreted by the fact that here the neutrino
eigenstate $\nu_3^m$ is heavier than $\nu_2^m$ (and in turn
$m_{\nu_2} > m_{\nu_1}$) since the massive neutral gaugino
[$\tilde B^0$, $\tilde W^0_3$] components of $\nu_3^m$ are larger
than in the $\nu_2^m$ state (in turn, than in $\nu_1^m$).

\section{Discussion and conclusions}
\label{conclusion}

In the context of the NMSSM with the presence of RPV couplings large enough to generate realistic neutrino masses,
a gravitino LSP of mass ${\cal O}(10^2)$ GeV -- as appears in supergravity models -- is
sufficiently stable from the point of view of the age of the universe.
\\ Nevertheless, the gravitino lifetime that we obtain is of order $\sim 10^{19}$ sec.
Now even in a supergravity situation with $m_{\rm 3/2} \lesssim 80$ GeV, where the opened decay channels
$\tilde{G}\rightarrow \gamma \nu^m$, $\tilde{G} \rightarrow \ell^{\pm} f \bar f'$ and
$\tilde{G} \rightarrow \nu^m f \bar f$ are thus reduced,
the total width looses typically four orders of magnitude only.
Hence, the HEAT excess in the positron fraction \cite{Koji} and the exotic positron source apparently detected by PAMELA \cite{Buch}
(which both require, in standard solutions, $m_{\rm 3/2} \sim 100-200$ GeV and $\tau_{\tilde{G}} \sim 10^{26}$ sec)
do not seem to be simultaneously explainable by the present RPV scenario reproducing the neutrino masses. The reason is that
the neutrino-neutralino mixing necessary to create large enough neutrino masses seems to induce too large gravitino RPV decay widths.
\\ Moreover, this result that the gravitino lifetime is systematically smaller than the lifetime needed to explain the PAMELA excess with a gravitino
dark matter also means that, in our dark matter scenario, the gravitino decays produce always too large fluxes which are excluded by
PAMELA in particular. Hence our conclusion is that a gravitino LSP cannot be a good dark matter candidate if there exist significant
RPV mixing terms (reproducing the neutrino masses).

In the case of the existence of a special supergravity scenario -- like e.g. some hybrid supergravity-GMSB models -- where
one could have $m_{\rm 3/2}$ below $\sim$ 10 GeV (in order to sufficiently reduce $\Gamma(\tilde{G}\rightarrow \gamma \nu^m)$)
and $M_1 \sim M_2 \sim 10^2$ GeV (suppressing the $\Gamma(\tilde{G} \rightarrow \ell^{\pm} f \bar f', \nu^m f \bar f)$ contribution \cite{recentCHOI}),
the gravitino lifetime would then be above the PAMELA limit of $\sim 10^{26}$ sec, making $\tilde{G}$ a stable and viable dark matter candidate
[with RPV couplings producing the neutrino masses].

Now if the supersymmetry breaking relies instead on a pure GMSB mechanism, the neutralino mass matrix includes new elements
and the gravitino mass can decrease typically to eV scale. Then, the gravitino is clearly a stable LSP since
we obtain a bound $m_{\rm 3/2} \lesssim$ 650 GeV -- 3 TeV from the requirement that its lifetime exceeds the age of the universe, if the
photino component of the neutrino [of value comparable to the above supergravity case] is sufficiently large to induce correct neutrino masses.
\\ In this GMSB case, the gravitino lifetime is $\sim 10^{56}$ sec. Hence the PAMELA data cannot be explained
with a gravitino dark matter.
However, the PAMELA flux excess can have e.g. alternative natural and astrophysical explanations, like in electron-positron pairs
produced by nearby pulsars \cite{Profumo}
\footnote{We also note that, anyway, the mentioned interpretation \cite{Koji} of the HEAT anomaly predicts an antiproton flux which tends to be too large,
although the prediction suffers from significant uncertainties. Besides, electron and positron fluxes from gravitino decays (whatever are the gravitino characteristics)
cannot explain \cite{Buch} both the PAMELA positron fraction and the electron plus positron flux measured by Fermi LAT \cite{LAT}.}.
By consequence, the GMSB framework could allow for the existence of a good gravitino dark matter candidate within the RPV-NMSSM.

Finally, we have checked numerically the robustness of the above conclusions on the gravitino stability
when extending to the realistic case of three neutrino flavors. To illustrate this, we have selected some
RPV-NMSSM parameter sets in agreement with the various constraints implemented in the NMHDECAY code
and reproducing the squared neutrino mass eigenvalue differences measured in oscillation experiments.

Concerning the stability of a sneutrino LSP with respect to $t_0$, within RPV versions of the NMSSM reproducing the wanted
neutrino mass scale, the result is negative. The lightest neutralino LSP is also not expected to be stable with respect to $t_0$.

Finally, we comment that all these difficulties one faces for having a viable dark matter candidate in RPV models are to be bridged
with the astrophysical problems one encounters in baryogenesis when RPV interactions are indeed turned on to generate
the neutrino masses \cite{baryo}.

\section*{Acknowledgments}

The authors thank P.~E.~Larre and Y.~Mambrini for useful discussions.
This work is supported by the European network HEPTOOLS and the A.N.R. {\it CPV-LFV-LHC} 
under project \textbf{NT09-508531} as well as 
the A.N.R. {\it TAPDMS} under project \textbf{09-JCJC-0146}.


\begin{thebibliography}{99}

\bibitem{SUGRA} D. Z. Freedman, P. van Nieuwenhuizen and S. Ferrara, Phys. Rev. D 13 (1976) 3214;
S. Deser and B. Zumino, Phys. Lett. B 62 (1976) 335.

\bibitem{SUGRAreport} H. P. Nilles, Phys. Rept. 110 (1984) 1.

\bibitem{GravDM} H. Pagels and J. R. Primack, Phys. Rev. Lett. 48 (1982) 223.

\bibitem{rpar1}
G. Farrar and P. Fayet, \Journal{\PLB}{76}{575}{1978};
S. Weinberg, \Journal{\PRD}{26}{287}{1982};
N. Sakai and T. Yanagida, \Journal{\NPB}{197}{533}{1982};
C. Aulakh and R. Mohapatra, \Journal{\PLB}{119}{136}{1982}.

\bibitem{Takayama} F. Takayama and M. Yamaguchi, Phys. Lett. B 485 (2000) 388.

\bibitem{NucleoLepto} W. Buchm\"uller, L. Covi, K. Hamaguchi, A. Ibarra and
T. Yanagida, JHEP 0703 (2007) 037.

\bibitem{Koji} K. Ishiwata, S. Matsumoto and T. Moroi, Phys. Rev. D 78 (2008) 063505;
A. Ibarra and D. Tran, JCAP 0807 (2008) 002.

\bibitem{HEAT} S. W. Barwick et al. [HEAT Collaboration], Astrophys. J. 482 (1997) L191.

\bibitem{PAMELA} O. Adriani et al. [PAMELA Collaboration], Nature 458 (2009) 607.

\bibitem{NMSSM}
H. P. Nilles et al., \Journal{\PLB}{120}{346}{1983};
J. M. Fr\`ere et al., \Journal{\NPB}{222}{11}{1983};
J. P. Derendinger and C. A. Savoy, \Journal{\NPB}{237}{307}{1984};
J. R. Ellis et al., \Journal{\PRD}{39}{844}{1989};
M. Drees, Int. J. Mod. Phys. A 4 (1989) 3635;
U. Ellwanger et al., \Journal{\PLB}{315}{331}{1993};
P. N. Pandita, Phys. Lett. B 318 (1993) 338;
P. N. Pandita, Z. Phys. C 59 (1993) 575;
S. F. King and P. L. White, \Journal{\PRD}{52}{4183}{1995};
F. Franke and H. Fraas, Int. J. Mod. Phys. A 12 (1997) 479;
M. Bastero-Gil et al., \Journal{\PLB}{489}{359}{2000}.

\bibitem{NMpheno}
M. Maniatis, Int. J. Mod. Phys. A 25 (2010) 3505;
U. Ellwanger, C. Hugonie and A. M. Teixeira, Phys. Rept. 496 (2010) 1.

\bibitem{AGG} A. Abada , G. Bhattacharyya  and G. Moreau, Phys. Lett. B 642 (2006) 503.

\bibitem{Valle} M. Maltoni, T. Schwetz, M. A. T\'ortola and J. W. F. Valle, New J. Phys. 6 (2004) 122;
see also the version 6 of {\tt arXiv:hep-ph/0405172}.

\bibitem{Gunion} R. Dermisek and J. F. Gunion, Phys. Rev. D 76 (2007) 095006;
U. Ellwanger, J. F. Gunion and C. Hugonie, JHEP 0507 (2005) 041;
  {\tt arXiv:hep-ph/0111179}.

\bibitem{ALEPH} K. Cranmer and P. Spagnolo, {\it Searching Higgs decaying to 4 taus}, Talk given at the Conference ``20 Years of ALEPH Data'', November 3, 2009, CERN,
available at {\tt http://indico.cern.ch/conferenceDisplay.py?confId=71475}

\bibitem{NMHDECAY} U. Ellwanger, J. F. Gunion and C. Hugonie, JHEP 0502 (2005) 066.

\bibitem{hybridI}
E. Poppitz and S. P. Trivedi, Phys. Rev. D 55 (1997) 5508 ; N.
Arkani-Hamed, J. March-Russell and H. Murayama, Nucl. Phys. B 509
(1998) 3; T. Gherghetta, G. F. Giudice and A. Riotto, Phys. Lett.
B 446 (1999) 28.

\bibitem{hybridII}
E. Dudas, Y. Mambrini, S. Pokorski and A. Romagnoni, JHEP 0804
(2008) 015; E. Dudas, Y. Mambrini, S. Pokorski, A. Romagnoni and
M. Trapletti, JHEP 0903 (2009) 011; D. G. Cerdeno, Y. Mambrini and A.
Romagnoni, JHEP 0911 (2009) 113.

\bibitem{thermal} W. Buchm\"uller, M. Endo and T. Shindou, JHEP 0811 (2008) 079.

\bibitem{Buch}
K. Ishiwata, S. Matsumoto and T. Moroi, {\tt arXiv:0811.0250 [hep-ph]};
JHEP 0905 (2009) 110; A. Ibarra and D. Tran, JCAP 0902 (2009) 021;
W. Buchm\"uller, A. Ibarra, T. Shindou, F. Takayama and D. Tran,
JCAP 0909 (2009) 021.

\bibitem{GMSB} G. F. Giudice and R. Rattazzi, Phys. Rept. 322 (1999) 419.

\bibitem{nmssmtools3}
  U. Ellwanger, C.-C. Jean-Louis and A. M. Teixeira,
  JHEP 0805 (2008) 044.

\bibitem{Flux} L. Covi, M. Grefe, A. Ibarra and D. Tran, JCAP 0901 (2009) 029.

\bibitem{FluxGamma} A. Ibarra and D. Tran, Phys. Rev. Lett. 100 (2008) 061301.

\bibitem{Moha} X. Ji, R. N. Mohapatra, S. Nussinov and Y. Zhang, Phys. Rev. D 78 (2008) 075032.

\bibitem{Pavel} P. F. Perez and S. Spinner, Phys. Lett. B 673 (2009) 251; {\tt arXiv:0909.1841 [hep-ph]};
V. Barger, P. F. Perez and S. Spinner, Phys. Rev. Lett. 102 (2009) 181802.

\bibitem{Endo} M. Endo and T. Shindou, JHEP 0909 (2009) 037.

\bibitem{rpNMSSM0} P. N. Pandita and P. F. Paulraj, Phys. Lett. B 462 (1999) 294;
P. N. Pandita, Phys. Rev. D 64 (2001) 056002.

\bibitem{rpNMSSM1} M. Chemtob and P. N. Pandita,
    \Journal{\PRD}{73}{055012}{2006};
     \Journal{\PRD}{76}{095019}{2007}.

\bibitem{rpNMSSM2} A. Abada and G. Moreau, JHEP 0608 (2006) 044.

\bibitem{SRoy} P. Ghosh and S. Roy, JHEP 0904 (2009) 069.

\bibitem{Munoz} D. E. Lopez-Fogliani and C. Mu\~{n}oz, Phys. Rev. Lett. 97 (2006) 041801.

\bibitem{Nilles} H. P. Nilles and N. Polonsky, Nucl. Phys. B 484 (1997) 33.

\bibitem{MunozBis} K. Y. Choi et al., {\tt arXiv:0906.3681 [hep-ph]}.

\bibitem{Dreiner} H. K. Dreiner and G. Moreau, Phys. Rev. D 67 (2003) 055005.

\bibitem{Gava}
  J. Gava and C.-C. Jean-Louis, Phys. Rev. D 81 (2010) 013003.

\bibitem{Moultaka}
S. Bailly, K. Jedamzik and G. Moultaka, Phys. Rev. D 80 (2009) 063509;
G. Moultaka, Acta Phys. Polon. B 38 (2007) 645.

\bibitem{PhysRep} R. Barbier et al., \Journal{\PR}{420}{1}{2005}.

\bibitem{nubounds} R. Hempfling, \Journal{\NPB}{478}{3}{1996}; E. J. Chun and
S. K. Kang, \Journal{\PRD}{61}{075012}{2000}; M. Hirsch et al.,
\Journal{\PRD}{62}{113008}{2000}; J. C. Romao, {\tt arXiv:hep-ph/0510411};
S. Rakshit, G. Bhattacharyya and A. Raychaudhuri,
\Journal{\PRD}{59}{091701}{1999}; G. Bhattacharyya,
H. V. Klapdor-Kleingrothaus and H. Pas, \Journal{\PLB}{463}{77}{1999};
S. Rakshit, Mod. Phys. Lett. A 19 (2004) 2239; Y. Grossman and
S. Rakshit, \Journal{\PRD}{69}{093002}{2004}.

\bibitem{REVbounds}
G. Bhattacharyya, Nucl. Phys. Proc. Suppl. A 52
(1997) 83, {\tt arXiv:hep-ph/9608415}; G. Bhattacharyya, Invited talk
presented at `Beyond the Desert', Castle Ringberg, Tegernsee, Germany,
8-14 June 1997, {\tt arXiv:hep-ph/9709395}; H. K. Dreiner, {\it
``Perspectives on Supersymmetry''}, Edited by G. L. Kane, World
Scientific, {\tt arXiv:hep-ph/9707435}; R. Barbier et al., {\tt
arXiv:hep-ph/9810232}; B. C. Allanach, A. Dedes and H. K. Dreiner,
\Journal{\PRD}{60}{075014}{1999};
\Journal{\PRD}{69}{115002}{2004}; Erratum-ibid. D 72 (2005) 079902;
M. Chemtob, Prog. Part. Nucl. Phys. 54 (2005) 71.

\bibitem{GravTri}
G. Moreau and M. Chemtob, Phys. Rev. D 65 (2002) 024033;
G. Moreau, Invited talk given at the XXXVIIth Rencontres de Moriond session devoted to
{\it ``Electroweak Interactions And Unified Theories''}, March 9-16 2002, Les Arcs (France),
{\tt arXiv:hep-ph/0205077}.

\bibitem{Hall} L. J. Hall and M. Suzuki, Nucl. Phys. B 231 (1984) 419.

\bibitem{PMNS} B. Pontecorvo, J. Exptl. Theoret. Phys. 33 (1957) 549; Sov. Phys. JETP 6 (1957) 429;
   Z. Maki, M. Nakagawa and S. Sakata, Prog. Theor. Phys. 28 (1962) 870;
   M. Kobayashi and T. Maskawa, Prog. Theor. Phys. 49 (1973) 652.

\bibitem{Lola} S. Lola, P. Osland and A. R. Raklev, Phys. Lett. B 656 (2007) 83.

\bibitem{ValleI} V. Berezinsky, A. Masiero and J. W. F. Valle, Phys. Lett. B 266 (1991) 382.

\bibitem{ValleII} V. Berezinsky, A. S. Joshipura and J. W. F. Valle, Phys. Rev. D 57 (1998) 147. 

\bibitem{Grefe} M. Grefe, PhD Thesis (2008), DESY-THESIS-2008-043.

\bibitem{Moroi} T. Moroi, PhD Thesis (1995), TU-479, {\tt arXiv:hep-ph/9503210}.

\bibitem{recentCHOI} K.-Y.~Choi and C.~E.~Yaguna, Phys. Rev. D 82 (2010) 015008.

\bibitem{cosmobound} S. Hannestad, \Journal {JCAP}{0305}{004}{2003};
K. Ichikawa, M. Fukugita and M. Kawasaki, \Journal{\PRD}{71}{043001}{2005};
J. Lesgourgues and S. Pastor, Phys. Rept. 429 (2006) 307.

\bibitem{absmass} Y. Farzan, O. L. G. Peres and A. Yu. Smirnov,
  \Journal{\NPB}{612}{59}{2001}; J. Bonn et al.,
  Nucl. Phys. Proc. Suppl. 91 (2001) 273; V. M. Lobashev
  et al., \Journal{\PLB}{460}{227}{1999};
  Nucl. Phys. Proc. Suppl. 77 (1999) 327; 91
  (2001) 280.

\bibitem{Profumo} S. Profumo, {\tt arXiv:0812.4457 [astro-ph]}.

\bibitem{LAT} A. A. Abdo et al. [The Fermi/LAT Collaboration], Phys. Rev. Lett. 102 (2009) 181101.

\bibitem{baryo} B.~A.~Campbell et al., Phys. Lett. B 256 (1991) 484; W.~Fischler et al., Phys. Lett. B 258 (1991) 45;
H.~K.~Dreiner and G.~G.~Ross, Nucl. Phys. B 410 (1993) 188.


\end{thebibliography}
\end{document}